\documentstyle[12pt]{article}
\title{Tetrad Gravity and Dirac's Observables.}
\author{\large Luca Lusanna \\[3mm]
\em Sezione INFN di Firenze, \\
\em Largo E.Fermi 2, 50125 Firenze, Italy\\
\em E-mail; lusanna@fi.infn.it}

\begin{document}
\maketitle

Talk given at the Conference ``Constrained Dynamics and Quantum
Gravity 99'', Villasimius (Sardinia, Italy), September 13-17, 1999.

\vskip 2cm

In a recent series of papers \cite{russo} the canonical reduction of a
new formulation of tetrad gravity to a canonical basis of Dirac's
observables in the 3-orthogonal gauge in the framework of
Dirac-Bergmann theory of constraints\cite{dirac} was studied.

This concludes the preliminary work in the research program aiming to
give a unified description of the four interactions in terms of
Dirac's observables. See Ref.\cite{india} for a complete review of the
achievements obtained till now:\hfill\break
 i) The understanding of the mathematical structures involved, in particular of the
Shanmugadhasan canonical transformations \cite{sha,lu1}.\hfill\break
 ii) The non-manifestly covariant canonical reduction to a canonical basis
 of Dirac's observables of many relativistic systems, including relativistic
 particles, the Nambu string, the electromagnetic, Yang-Mills and Dirac fields,
 the standard SU(3)xSU(2) xU(1) model of elementary particles in Minkowski
 spacetime. In the case of gauge theories, this required an understanding of all
 the pathologies of the constraint manifold associated with the Gribov ambiguity (gauge
 symmetries and gauge copies) and of the fact that the presence or absence of the
 Gribov ambiguity depends on the choice of the function space for the gauge fields
 and the gauge transformations. With the hypothesis that no physics is hidden in
 the Gribov ambiguity, on can work in special weighted Sobolev spaces \cite{moncr}
 where it is absent.
 Then, in the case of trivial principal bundles on constant time hypersurfaces in
 Minkowski spacetime (no monopoles; winding number but not instanton number) and
 for suitable Hamiltonian boundary conditions on gauge potentials and gauge
 transformations (the behaviour at spatial infinity must be direction-independent)
 allowing the color charges, in the case of SU(3), to be well defined, one can do
 a complete canonical reduction of Yang-Mills theory like in the electromagnetic
 case and find the singularity-free physical Hamiltonian.\hfill\break
 iii) The definition of the Wigner-covariant rest-frame instant form of dynamics
 (replacing the non-relativistic separation of the center-of-mass motion) for the
 timelike configurations of every isolated relativistic system
 (particles, strings, fields) in Minkowski spacetime. This is obtained starting
 from the reformulation of the isolated system on arbitrary spacelike
 hypersurfaces (parametrized Minkowski theories) and making a restriction
 to the special foliation (3+1 splitting) of Minkowski spacetime with Wigner
 hyperplanes: they are determined by the given configuration of the
 isolated system, being orthogonal to its conserved total 4-momentum (when
 it is timelike). A general study of the relativistic center of mass, of the
 rotational kinematics and of Dixon multipolar expansions \cite{dixon} is now
 under investigation \cite{iten} for N-body systems. See Refs.\cite{mate,lon}
 for the center of mass of a Klein-Gordon configuration.\hfill\break
 iv) The Wigner-covariant reformulation of the previous canonical reductions in the
 rest-frame instant form taking into account the stratification of the
 constraint manifold associated with the isolated system induced by the
 classification of its configurations according to the Poincar\'e orbits
 for the total 4-momentum.\hfill\break
 v) The realization that in the rest-frame instant form there is a universal
 breaking of Lorentz covariance regarding only the decoupled canonical
 non-covariant ``external" center of mass (the classical analogue of the
 Newton-Wigner position operator), while all the relative degrees of freedom are
 Wigner-covariant. The spacetime spreading of this non-covariance determines
 a classical unit of length, the M\o ller radius\cite{mol}, which is determined
 by the value of the Poincar\'e Casimirs of the given configuration of the
 isolated system and which should be used as a physical ultraviolet cutoff
 in quantization. The M\o ller radius is a non-local effect of Lorentz
 signature: already at the classical level it is impossible to localize
 in a covariant way the canonical center of mass of an isolated extended
 relativistic system with a precision better of this radius. This classical
 problem happens at those distances where quantum mechanics introduces pair
 creation (the M\o ller radius is of the order of the Compton wavelength of the isolated
 system). Moreover, the M\o ller radius is a remnant in Minkowski spacetime of
 the energy conditions of general relativity. With the methods of Ref.\cite{pauri}
 one can find the ``internal" 3-center of mass inside the Wigner hyperplane,
 whose vanishing is the gauge fixing for the constraints defining the rest frame.\hfill\break
 vi) Since the rest-frame instant form is a special classical background for the
 Tomonaga-Schwinger formulation of quantum field theory, there is now the possibility
 to start with a Wigner-covariant quantization of field theory on Wigner hyperplanes.
 Having built-in a covariant concept of ``equal time", one expects to find a
 Schroedinger-like equation for relativistic bound states (avoiding the problem
 of the spurious  solutions of the Bethe-Salpeter equation), to be able to define
 Tomonaga-Schwinger asymptotic states (with the possibility of including
 bound states among them) and to use the M\o ller radius as a physical ultraviolet
 cutoff.

The next conceptual problem was to apply all the technology developed
for constrained systems in Minkowski spacetime to a formulation of
general relativity able to incorporate the standard model of
elementary particles and such that it could be possible to formulate a
deparametrization scheme according to which the switching off of the
Newton constant reproduces the description of the standard model on
the Wigner hyperplanes in Minkowski spacetime. In this way, at least
at the classical level, the four interactions would be described in a
unified way and one could begin to think how to make their
quantization in a way avoiding the existing incompatibilty between
quantum mechanics and general relativity.

Tetrad gravity, rather than metric gravity, was the natural
formulation to be used for achieving this task for the following
reasons:\hfill\break
 i) The fermions of the standard model couple naturally to tetrad gravity.\hfill\break
 ii) Tetrad gravity incorporates by definition the possibility to have the matter
 described by an arbitrary (geodesic or non-geodesic) timelike congruence of observers.
 In this way one can arrive at a Hamiltonian treatment of the precessional aspects of
 gravitomagnetism like the Lense-Thirring effect\cite{ciuf}.\hfill\break
 iii) In tetrad gravity it is possible to replace the supermomentum constraints with
 SO(3) Yang-Mills Gauss laws associated with the spin connection and to solve them with the
 technology developed for the canonical reduction of Yang-Mills theories. Instead
 in metric gravity one does not know how to solve the supermomentum constraints.

 Let us remark that till now supergravity and string theories have not been analyzed,
 since the emphasis is on learning how to make the canonical reduction in presence
 of constraints and from this point of view these theories only have bigger gauge
 groups and many more constraints to be solved.

Another important point is that the dominant role of the Poincar\'e
group and of its representations in the theory of elementary particles
in Minkowski spacetime requires to formulate general relativity on
non-compact spacetimes asymptotically flat at spatial infinity so that
the asymptotic Poincar\'e charges \cite{adm,reg} exist and are well
defined. In presence of matter these asymptotic Poincar\'e charges
must reproduce the ten conserved Poincar\'e generators of the isolated
system with same matter content when the Newton constant is switched
off.

All these requirements select a class of spacetimes with the following
properties:\hfill\break
 i) They are pseudo-Riemannian globally
hyperbolic 4-manifolds $(M^4\approx R\times \Sigma , {}^4g)$ [$(\tau
,\vec \sigma ) \mapsto z^{\mu}(\tau ,\vec \sigma )$]. These spacetimes
have a global time function $\tau (z)$ and admit 3+1 splittings
corresponding to foliations with spacelike hypersurfaces
$\Sigma_{\tau}$ (simultaneity 3-manifolds, which are also Cauchy
surfaces).\hfill\break
 ii) They are non-compact and asymptotically flat at spatial infinity.\hfill\break
  iii) They are parallelizable 4-manifolds, namely they admit a spinor structure and have
  trivial orthonormal frame principal SO(3)-bundles over each simultaneity 3-manifold
  $\Sigma_{\tau}$.\hfill\break
  iv) The non-compact parallelizable simultaneity 3-manifolds $\Sigma_{\tau}$ are assumed to
  be topologically trivial, geodesically complete and diffeomorphic to $R^3$
  [$\Sigma_{\tau}\approx R^3$]. This implies the existence of global coordinate systems
  on $\Sigma_{\tau}$, so that coordinate systems $(\tau ,\vec \sigma )$, adapted to the
  simultaneity 3-surfaces $\Sigma_{\tau}$, can be used for $M^4$.
  In this simplified case the geodesic exponential map is
  a diffeomorphism, there are no closed 3-geodesics and no conjugate Jacobi
  points on 3-geodesics. \hfill\break
  v) The cotriads on $\Sigma_{\tau}$ and the associated 3-spin-connection
  on the orthogonal frame SO(3)-bundle over $\Sigma_{\tau}$ are assumed to belong
  to suitable weighted Sobolev spaces so that the Gribov ambiguity is absent. This implies
  the absence of isometries (and of the associated Killing vectors) of the non-compact
  Riemannian 3-manifold $(\Sigma_{\tau}, {}^3g)$.\hfill\break
  vi) Diffeomorphisms on $\Sigma_{\tau}$ and their extension to tensors are interpreted
  in the passive sense (pseudo-diffeomorphisms), following Ref.\cite{be}, in accord
  with the Hamiltonian point of view that infinitesimal diffeomorphisms on tensors are
  generated by taking the Poisson bracket with the first class supermomentum constraints.

As action principle we use the ADM metric action with the 4-metric
${}^4g$ rewritten in terms of general cotetrads on $M^4$. For the
general cotetrads a new special parametrization has been found.
Starting from $\Sigma_{\tau}$-adapted cotetrads (Schwinger time gauge)
whose 13 degrees of freedom are the lapse and shift functions and the
cotriads on $\Sigma_{\tau}$, the remaining 3 degrees of freedom are
described by the 3 parameters which parametrize timelike Wigner boosts
acting on the flat indices of the cotetrad (in the cotangent spaces
over each point of $\Sigma_{\tau}$). This implies that the flat
indices acquire Wigner covariance (the time index becomes a Lorentz
scalar, while the spatial indices become Wigner spin 1 3-indices) in
each point of $\Sigma_{\tau}$. These 3 boost parameters describe the
transition from the $\Sigma_{\tau}$-adapted Eulerian observers
associated with the $\Sigma_{\tau}$-adapted tetrads (this timelike
congruence is surface-forming and is orthogonal to the
$\Sigma_{\tau}$'s) to an arbitrary (in general not surface-forming)
timelike congruence of observers.

The ADM Lagrangian density is considered a function of these 16
fields: the lapse and shift functions, the cotriads on
$\Sigma_{\tau}$, the 3 boost parameters. In tetrad gravity there are
14 first class constraints (10 primary and 4 secondary
ones):\hfill\break
 i) The momenta conjugate to the lapse and shift functions vanish, so that lapse and shifts
 are 4 arbitrary gauge variables [arbitrariness in the choice of the standard of proper time
 and conventionality in the choice of the notion of simultaneity with the associated
 possible anisotropy in light propagation].\hfill\break
 ii) The momenta conjugate to the 3 boost parameters vanish (Abelianization
 of the Lorentz boost contained in the SO(3,1) group acting on the
 flat indices of the cotetrads): the 3 boost parameters are arbitrary gauge variables
 (the physics does not depend on the choice of the timelike congruence of
 observers).\hfill\break
 iii) There are 3 constraints describing the generators of SO(3) rotations on
 the flat indices of the cotetrads: the associated 3 angles (3 degrees of freedom
 among the 9 parametrizing the cotriads) are gauge variables (conventionality in the
 choice of the standard of non-rotation for a timelike congruence of observers).
 \hfill\break
 iv) There are 3 secondary constraints which are equivalent to the ADM supermomentum
 ones (it is possible to replace them with SO(3) Yang-Mills Gauss laws for the
 spin connection): 3 degrees of freedom, depending on the cotriads and their
 time derivatives, are arbitrary gauge variables describing the freedom in the
 choice of the 3-coordinates on $\Sigma_{\tau}$ (arbitrariness in the choice of
 3 standards of length). These 3 constraints generate the pseudo-diffeomorphisms.
 \hfill\break
 v) One secondary constraint coincides with the ADM superhamiltonian one. It can be shown
that this constraint has to be interpreted as the Lichnerowicz
equation \cite{conf}determining the conformal factor of the 3-metric
${}^3g$. Therefore, the last gauge variable is the momentum conjugate
to this conformal factor [it is non-locally connected with the trace
of the extrinsic curvature of $\Sigma_{\tau}$, also named York time
\cite{yoyo}] and the gauge transformations generated by the
superhamiltonian constraint correspond to the transition from one
allowed 3+1 splitting of $M^4$ with spacelike hypersurfaces
$\Sigma_{\tau}$ to another one (the physics does not depend on the
choice of the 3+1 splitting, like in parametrized Minkowski theories).

In conclusion, there are only two dynamical degrees of freedom hidden
in the cotriads on $\Sigma_{\tau}$ and they describe the gravitational
field. Their determination requires a complete breaking of general
covariance, namely a complete fixation of the gauge degrees of freedom
(this amounts to the choice of a physical laboratory where to do all
the measurements).

Let us remark that the fixation of the 3-coordinates and of the 3
rotation angles are inter-related, because the associated constraints
do not have vanishing Poisson brackets. Moreover, there are
restrictions on the gauge transformations when one restricts himself
to the solutions of Einstein's equations: according to the general
theory of constraints one has to start by adding the gauge fixings to
the secondary constraints; the requirement of their time constancy
generates the gauge fixings of the primary constraints. Therefore,
since the supermomentum constraints are secondary ones, the choice of
the 3-coordinates on $\Sigma_{\tau}$ determines the choice of the
shift functions (i.e. of the associated convention for simultaneity in
$M^4$; the Einstein convention can be applied only when the shift
functions vanish). Analogously, the choice of the 3+1  splitting of
$M^4$ (fixation of the momentum conjugate to the conformal  factor of
the 3-metric) determines the choice of the lapse function (namely of
how the 3-surfaces $\Sigma_{\tau}$ are packed in the chosen 3+1
splitting of $M^4$).

The next problem is the choice of the boundary conditions for the 16
fields in the cotetrads and for the allowed Hamiltonian gauge
transformations. The existence of the Poisson brackets and the
differentiability of the Dirac Hamiltonian require the addition of a
surface term \cite{witt} to the Dirac Hamiltonian containing the
strong ADM Poincar\'e charges: they are surface integrals at spatial
infinity, which differ from the weak ADM Poincar\'e charges (volume
integrals) by terms vanishing due to the secondary constraints. In
spacetimes asymptotically flat at spatial infinity besides the 10
asymptotic Poincar\'e charges there is a double infinity of Abelian
supertranslations (associated with the asymptotic direction-dependent
symmetries of these spacetimes \cite{ash}). Their presence generates
an infinite-dimensional algebra of asymptotic charges which contains
an infinite number of conjugate Poincar\'e subalgebras: this forbids
the identification of a well defined angular momentum in general
relativity. The requirement of absence of supertranslations, so to
have a uniquely defined asymptotic Poincar\'e algebra, puts severe
restrictions on the boundary conditions of the 16 fields and of the
gauge transformations.

Following Dirac \cite{dirac}, we assume the existence of asymptotic
flat coordinates for $M^4$. It can be shown that this implies the
restriction of the allowed 3+1 splittings of $M^4$ to those whose
associated foliations have the leaves $\Sigma_{\tau}$ approaching
spacelike Minkowski hyperplanes at spatial infinity. The absence of
supertranslations requires that this approach must happen in a
direction-independent way and that the lapse and shift functions can
be consistently written as an asymptotic part (equal to the lapse and
shifts of spacelike Minkowski hyperplanes) plus a part which vanishes
at spatial infinity. Since spacelike Minkowski hyperplanes are
described in phase space by 10 configuration variables (an origin plus
an orthonormal tetrad) plus the conjugate momenta (see the
parametrized Minkowski theories), Dirac adds these 20 variables to the
ADM phase space, but then he also adds 10 first class constraints to
the Dirac Hamiltonian (so that the 10 configurational variables are
gauge variables). These constraints determine the 10 extra momenta in
terms of the 10 weak Poincar\'e charges.

The satisfaction of all the requirements on the boundary conditions of
the 16 fields and of the gauge transformations, in particular the
absence of supertranslations, leads to the following results. The
Hamiltonian formulation of both metric and tetrad gravity is well
posed for the class of Christodoulou-Klainermann spacetimes \cite{ck},
which are near Minkowski spacetime in a norm sense and avoid the
singularity theorems not admitting a conformal completion, but which
contain asymptotic gravitational radiation at null infinity (even if
with a weaker peeling of the Weyl tensor). The allowed 3+1 splittings
for these spacetimes have all the leaves $\Sigma_{\tau}$ approaching,
in a direction-independent way, those special Minkowski hyperplanes
asymptotically orthogonal to the weak ADM Poincar\'e 4-momentum. These
asymptotic spacelike hyperplanes are the analogue of the Wigner
hyperplanes of parametrized Minkowski theories, and, when matter is
present, allow to deparametrize tetrad gravity so to obtain the
description of the same matter in the rest-frame instant form on
Wigner hyperplanes in Minkowski spacetime when the Newton constant is
switched off. Therefore, this Hamiltonian treatment of the
Christodoulou-Klainermann spacetimes is the rest-frame instant form of
general relativity; like in parametrized Minkowski theories, there is
a decoupled canonical non-covariant ``external" center of mass (a
point particle clock) now located near spatial infinity, while all the
physical degrees of freedom are relative variables (a weak form of
Mach principle). These asymptotic hyperplanes are privileged observers
dynamically selected by the given configuration of the gravitational
field (they replace the ``fixed stars") and not a priori given like in
bimetric theories or in theories with a background. It can be shown
that given an asymptotic tetrad determined by the ADM 4-momentum, this
tetrad can be transported in each point of $\Sigma_{\tau}$ by using
the Frauendiener equations\cite{p28}  with the Sen connection
(replacing the Sen-Witten equations \cite{p27} for spinors in the case
of triads and tetrads), so determining a dynamically selected
privileged timelike congruence of observers. These spacelike
hypersurfaces $\Sigma_{\tau}$ can be called Wigner-Sen-Witten (WSW)
hypersurfaces.

Given this framework, it is possible to solve the rotation and
supermomentum constraints and to find  parametrizations of the
cotriads in terms of:\hfill\break
 i) the 3 gauge rotation angles;\hfill\break
 ii) the 3 gauge parameters associated with the pseudodiffeomorphisms,
 namely with the choice of the 3-coordinates;\hfill\break
 iii) the conformal factor of the 3-metric;\hfill\break
 iv) the two physical degrees of freedom describing the gravitational field.
 \hfill\break
 Each choice of the 3-coordinates on $\Sigma_{\tau}$ turns out to be equivalent
 to the choice of a particular parametrization of the cotriad
 (see Refs.\cite{cji} for previous attempts).
 In this way 13 of the 14 first class constraints are under control and we can do
 a Shanmugadhasan canonical transformation adapted to these 13 constraints.

We have till now studied the most natural choice of 3-coordinates,
which corresponds to the 3-orthogonal gauges in which the 3-metric is
diagonal (they are the nearest ones to the standards of the
non-inertial physical laboratories on the earth). The 3 rotation
angles and the 3 boost parameters are put equal to zero. As a last
gauge fixing we put equal to zero the momentum conjugate to the
conformal factor of the 3-metric, which, in turn, must be determined
as solution of the Lichnerowicz equation in this gauge. By going to
Dirac brackets with respect to the 14 constraints and the 14
gauge-fixings, we remain with two pairs of canonical variables ,
describing the Hamiltonian physical degrees of freedom or Dirac's
observables of the gravitational field in this completely fixed
3-orthogonakl gauge (complete breaking of general covariance). The
physical Hamiltonian for the evolution in the mathematical time
parametrizing the WSW hypersurfaces $\Sigma_{\tau}$ of the foliation
is the weak (volume form) ADM energy \cite{cho}: it depends only on
the Dirac's observables, even if part of the dependence is through the
conformal factor of the 3-metric, whose form is unknown since noone is
able to solve the Lichnerowicz equation. The physical times (atomic
clocks, ephemeris times,..) have to be locally correlated to this
mathematical time.

Also the Komar-Bergmann individuating fields \cite{komar}, needed for
a physical identification of the points of the spacetime $M^4$ (due to
general covariance the mathematical points of $M^4$ have no physical
meaning in absence of a background; see Einstein's hole argument), may
be re-expressed in terms of Dirac's observables.

Finally the Poincar\'e Casimirs associated with the asymptotic weak
Poincar\'e charges allow to define the M\o ller radius (and a possible
ultraviolet cutoff in a future attempt to make a quantization of
completely gauge-fixed tetrad gravity) also for the gravitational
field.

The main tasks for the future are:

A) Make the canonical quantization of scalar electrodynamics in the
rest-frame instant form on the Wigner hyperplanes, which should lead
to a particular realization of Tomonaga-Schwinger quantum field
theory, avoiding the no-go theorems of Refs.\cite{torre}. The M\o ller
radius should be used as a physical ultraviolet cutoff for the point
splitting technique and the results of Refs.\cite{lavelle} about the
infrared dressing of asymptotic states in S matrix theory should help
to avoid the `infraparticle' problem\cite{buc}.

B) Study the linearization of tetrad gravity in the 3-orthogonal gauge
to reformulate the theory of gravitational waves in this gauge.

C) Study the N-body problem in tetrad gravity at the lowest order in
the Newton constant (action-at-a-distance plus linearized tetrad
gravity). See Ref.\cite{russo} for preliminary results on the action
at a distance hidden in Einstein's theory at the lowest order in the
Newton constant, which agree with the old results of Ref.\cite{dro}.

D) Study the perfect fluids both in the rest-frame instant form in
Minkowski spacetime \cite{milla} and in tetrad gravity.

E) Make the Hamiltonian reformulation of the Newman-Penrose
formalism\cite{stew} by using Hamiltonian null tetrads and study its
connection with the 2+2 decompositions of $M^4$\cite{inverno}.

F) Begin the study of the standard model of elementary particles
coupled to tetrad gravity starting from the Einstein-Maxwell system.

\end{document}